\documentclass[a4paper,12pt]{article}
\usepackage{amssymb,amsmath,amsthm,mathrsfs,mathtools}
\usepackage{caption}
\usepackage{cite}
\usepackage{hyperref}
\usepackage[affil-it]{authblk}
\usepackage{tikz,pgfplots,tkz-euclide}
\usetikzlibrary{arrows,positioning,calc,fadings,decorations.pathreplacing,decorations}
\def\R{\mathbf{R}}
\newcommand\C{\mathbf{C}}
\def\H{\mathbf{H}}

\def\P{\mathbf{P}}
\def\q{\mathbf{q}}
\newcommand\G[1]{\Gamma(#1)}

\newcommand\eq[1]{(\ref{#1})}

\def\pa{\partial}

\def\conf{{\mathcal{C}}}

\def\z{{\boldsymbol{z}}}
\def\m{{\boldsymbol\mu}}
\def\bxi{{\boldsymbol\xi}}
\newcommand\inmu[3]{I^{{#2}}_{{#1}}({#3})}
\def\inmuz{\inmu{N}{\m}{{\z}}}
\newcommand\rt{\longrightarrow}

\def\I{\mathcal{I}}

\def\Pid{\Pi_{\Delta}}
\newcommand\norm[1]{\left\Vert #1\right\Vert}
\def\Q{{\boldsymbol Q}}
\def\Qd{Q^{\dagger}}
\def\tr{\text{Tr\ }}

\def\vt{\vartheta}
\title{Conformal Correlation functions in four dimensions from 
Quaternionic Lauricella system}
\author{Aritra Pal\thanks{email: intap@iacs.res.in}~} 
\author{Koushik Ray\thanks{email: koushik@iacs.res.in}}
\affil{Indian Association for the Cultivation of Science,\authorcr
Calcutta 700 032. India.}
\date{}
\begin{document}
\maketitle
\begin{abstract}
\noindent
Correlation functions in four-dimensional Euclidean conformal field theories 
are expressed in terms of representations of the conformal group 
$SL(2,\H)$, $\H$ being the field of quaternions, on the configuration space
of points. The representations are obtained in terms of a Lauricella system
derived using quaternions. It generalizes the two-dimensional case, 
wherein the $N$-point correlation function is
expressed in terms of solutions of Lauricella system on the configuration space
of $N$ points on the complex plane, furnishing representation of the 
conformal group $SL(2,\C)$.
\end{abstract}
\setcounter{page}{0}
\thispagestyle{empty}
\clearpage
\section{Introduction}
Correlation functions in conformal field theories in
various dimensions have been studied extensively. 
Recent impetus to this field came from the conformal
bootstrap programme \cite{epfl,tasi,qual,prv}. 
Correlation functions of conformal fields 
at different points in a geometric space are obtained as equivariant 
quantities under the conformal group of the space. 
That is, correlation functions are appropriate representations of the 
conformal group. A representation of a
group acting on  a topological space is given by the lift of the group action
to the space of regular functions on the topological space, or their
appropriate generalizations. If the topological space is non-compact,
functions on some form of completion of it is considered in order to ensure
convergence of various functions and integrals. 
For conformal groups it is customary to use a conformal compactification. 
From now on we shall restrict 
our discussion to the $n$-dimensional Euclidean spaces, $\R^n$. 
In this case a popular scheme is to 
consider the action of the conformal group of $\R^n$ isomorphic to
$SO(1,n+1)$ on the light cone of
$\R^{n+2}$ with a metric of signature $(-,+,+,\cdots)$. 
The light cone is stabilized by the conformal group. The Euclidean space 
$\R^n$ is embedded into the light cone by an injective map. Its completion to 
include the conformal infinity is then used to construct representations of
the conformal group. 
For example, in order to
obtain the correlation functions of conformal fields on the complex plane
$\C$, one first obtains the representation of the global conformal
group $SO(1,3)$ or  $SL(2,\C)$, on the conformal compactification 
of $\C$, namely, $\P^1$, the
complex projective line embedded into the light cone in $\R^4$. 
Functions on the completion obtained by restriction from the
light cone in two higher dimensions 
are acted on by the conformal group, thereby furnishing its 
representation. This picture, however, pertains to a single field in 
$\R^n$. Correlation functions for a multitude of 
fields are obtained by tensoring such representations. 
The correlation functions are then
arranged into conformal blocks, the eigenfunctions of the quadratic Casimir,
expanded in the basis of asymptotic plane waves. 
Since the conformal group includes scaling, construction of such a
representation is often facilitated by considering the Mellin 
transforms \cite{mack,gkss,gkss0,pen,pao1,pao2}. 
While the two-point and
three-point functions are determined by the conformal group and the structure
constants, higher point correlation functions require further
restrictions to be imposed. The bootstrap constraint, which has been a topic
of extensive discussion recently, is one such \cite{polyakov,gs}, which
restricts the correlation functions by its properties under the permutation
of the points. 

The representations, equivariant as they are, do not capture the 
nuances of various conformal field theories. These are incorporated by
inserting projectors in the correlation functions such that higher point 
functions are expressed in terms of three-point functions. 
The projectors
are made up of fields in a specific field theory. Hence the three point
functions carry the structure constants of the operators of the same theory.
We shall make extensive use of this formalism, called the shadow operator
formalism \cite{fgpg,do1,sd}.

In this article we obtain the multi-point correlation functions of conformal
field theories in two- and four-dimensional Euclidean spaces in terms of 
representations of the corresponding conformal groups. Instead of tensoring 
the ``single-particle" 
representations of the Lie algebra of the conformal group,
we approach the computation of $N$-point correlation
functions by looking at the representation of the M\"obius
group on the configuration space of $N$ marked points on the Euclidean space. 
Among the various models of the configuration space the one we use is
the Fulton-Macpherson compactification of the space of $N$ 
pairwise distinct points.

In two dimensions we consider $N$ points on 
the complex plane $\C$. The representation of the   
conformal group $SL(2,\C)$ is then sought among the germ of 
functions, described by a
Lauricella system, on the configuration space. The Lauricella system 
is given by the solutions of a system of differential equations in terms of
the positions of the $N$ points. The correlation functions are furnished by 
the ones equivariant under $SL(2,\C)$. At this level, the completion 
of the configuration space is
brought about by demanding that the functions are regular
at infinity. 
The two-dimensional conformal group generalizes to the M\"obius
group $SL(2,\H)$ in four dimensions, where $\H$ denotes the field of 
quaternions \cite{wilker,porteous,baez}.
We show that the Lauricella system has an
appropriate generalization in terms of quaternionic variables. The correlation
functions are once again given by the equivariant ones, regular at infinity.
In both cases we deal with the conformal group, rather than the algebra. 
Higher point functions are split using the projectors and related to
integrals over the $N$-variable Lauricella functions, dispensing with
the point-wise insertion of ``single-particle" Casimirs which proved to
be useful too \cite{do1,do,rosenhaus,parikh,fs,fms1}. 
The integrals involved in the correlation functions appeared earlier
literature \cite{Sym,DF,do,elkhidir,rrtv,lmm}. These are similar to Feynman integrals in higher
dimensions. However, direct evaluation of the integrals is rendered difficult
by their multi-valued nature and is greatly facilitated by writing them
as solution to differential equations. 
We find that the differential
equations of the Lauricella system have a close 
analogue in four dimensions in terms of matrix-valued quaternions. 
The equation for the
general case with an arbitrary number of points has been written
down.  

In the next section we describe the Lauricella system on the configuration
space of marked points in the two-dimensional case \cite{yoshida,looi} 
and their appearance in the computation of chiral
correlation functions through representation of the
M\"obius group. The projector is given by a two-point Lauricella function too. 
We explicitly evaluate the four and five point integrals and express the
corresponding correlation functions in terms of integrals involving them,
reproducing previously known results, as expected.
The four-point function is expressed in terms of the Gauss hypergeometric 
function, while the five point function is expressed in terms of the Appell
function $F_2$. In the third section 
generalization to four dimensions is carried out.
First, the complex integrals are generalized to integrals over quaternions, 
which generalize the field-theoretic  Feynman integrals in four-vectors. By
taking derivatives with respect to the matrix-valued quaternions we then obtain
differential equations generalizing the Lauricella system to four dimensions.
Let us stress that while the integrals appearing in the correlation
functions have long been known \cite{Sym} as integrals over
four-vectors, Lauricella-type differential equations to evaluate them, to the
best of our knowledge, have not appeared earlier. Let us also point out that 
the multi-valued integrals are expressed in terms of linear combinations
solutions of the Lauricella system. As has been experienced in the evaluation
of period integrals in the studies of mirror symmetry, obtaining them as
solutions to differential equations may be more efficient for the evaluation
of the integrals compared to direct computation.
We then show that these integrals furnish representation of the four-dimensional
M\"obius group $SL(2,\H)$ by enumerating their transformation under the 
group.
Equations for the invariant part of the integrals, 
which may be related to the conformal block, expressed in terms of 
cross-ratios defined as determinants of a product of a quartet of quaternions 
and then obtained by taking traces of the matrix equations. We present the
results for the case of four points, where the Lauricella system is solved
with the Appell function $F_4$. 
\section{Two dimensions}
\subsection{Functions on the configuration space of points}
Let us begin with a description of the functions on the 
configuration space of $N$ distinct
points $\{z_1,z_2,\cdots,z_N\}$ on the complex plane $\C$. The configuration
space is 
\begin{equation}
\conf_N(\C) =\C^N\setminus\cup_{1\leq i,j\leq N} \Delta_{ij},
\end{equation} 
where 
\begin{equation} 
\Delta_{ij}=\{(z_1,z_2,\cdots,z_N)\in\C^N;z_i = z_j \}
\end{equation}  
is called the fat diagonal. 
On the configuration space one considers integrals of the form
\begin{equation}
\label{IN}
I^{\m}_N(\z)=\int\frac{dz}{(z-z_1)^{\mu_1}
(z-z_2)^{\mu_2}\cdots
(z-z_N)^{\mu_N}},
\end{equation} 
where vectors in boldface denote the $N$-tuples.
The vector $\z=(z_1,z_2,\cdots,z_N)$ collects the positions of the $N$ points
and $\m=(\mu_1,\mu_2,\cdots,\mu_N)$ is the $N$-tuple of parameters, called
weights. The integral is defined over an arc in the plane connecting a pair of 
zeroes of the denominator of the integral, avoiding encircling any other zero
and $0<\mu_i<1$ for each $i=1,2,\cdots,N$. This integral defines a local
system of $\C$-vector spaces over $\conf_N(\C)$, whose stalk at a point
$\z$ will also be
denoted $\inmuz$ by abuse of notation. Then $\inmuz$ is invariant under translation
of $\z$ by a constant, is homogeneous of degree $1-|{\boldsymbol\mu}|$, where 
$|{\boldsymbol\mu}|=\sum_{i=1}^N\mu_i$, and
satisfies the differential equation \cite{looi}
\begin{equation}
\label{laueq}
z_{ij}\frac{\pa^2\inmuz}{\pa z_i\pa z_j} = 
\mu_j\frac{\pa\inmuz}{\pa z_i}
-\mu_i\frac{\pa\inmuz}{\pa z_j},
\end{equation} 
where we used $z_{ij}=z_i-z_j$.
This equation is obtained by differentiating \eq{IN}
with respect to the $z_i$ under the integral sign and using the identity
\begin{equation}
	\label{identity3}
	\frac{1}{(x-y)(y-z)}+\frac{1}{(y-z)(z-x)}+\frac{1}{(z-x)(x-y)}=0
\end{equation}
of three complex numbers $x$, $y$, $z$.
The germs of $\inmuz$ 
are expressed as the germs of the Lauricella functions
\cite{looi}, determined uniquely by \eq{laueq}. 
We refer to equation \eq{laueq} and its solutions  as the
Lauricella system.
In mundane terms, the solutions of  equation \eq{laueq} are ``good
functions" on the completion of the configuration space $\conf_N(\C)$.

Invariance under translation by a constant implies that $\inmuz$ depends only on the 
differences $z_{ij}$ and
not separately on $z_i$ themselves. The integral is well-behaved at infinity provided 
$|\m|=2$, as can be 
checked by changing the integration variable $z$ to $1/z$. The case of
$N=2$ require special 
treatment. Let us discuss it first. 
Since the integral involves only two marked points, $z_1$ and
$z_2$, we can take the path over any arc joining these two points, 
which is in fact homotopic to  the line joining them. Thus,
\begin{equation}
	\inmu{2}{(\mu_1,\mu_2)}{z_1,z_2}=\int_{z_1}^{z_2}\frac{dz}{(z-z_1)^{\mu_1}(z-z_2)^{\mu_2}}.
\end{equation}
Parametrizing the line joining the two points as 
$z=tz_2+(1-t)z_1$, such that $0\leq t\leq 1$, the integral
is evaluated to be 
\begin{equation}
	\label{I2:gen}
	\inmu{2}{(\mu_1,\mu_2)}{z_1,z_2} = \frac{1}{z_{12}^{\mu_1+\mu_2-1}}
	\frac{\G{1-\mu_1}\G{1-\mu_2}}{\G{2-\mu_1-\mu_2}}.
\end{equation}
Here and in the following we ignore factors of powers 
of $-1$, which can be absorbed in the 
normalization of the correlation functions. As mentioned before, 
the integral depends only on the difference $z_{12}$ 
rather than on the points individually and is homogeneous of degree
$1-\mu_1-\mu_2$.
The integral is, on the other hand, not well-behaved at infinity 
unless $\mu_1+\mu_2=2$, a feature 
to be called on later. When $\mu_1+\mu_2=2$, it becomes 
\begin{equation}
	\label{I2m2}
	\inmu{2}{(\mu_1,\mu_2)}{z_1,z_2} = \frac{1}{z_{12}}\frac{\G{1-\mu_1}\G{\mu_1-1}}{\G{0}},
\end{equation}
where the singular piece $\G{0}$ is to be understood in a limiting sense. 
Demanding the integrals to be regular at infinity is equivalent to considering
a completion of the configurations space. We work with
the Fulton-Macpherson compactification \cite{sinha,kmv} as discussed in
section \ref{sec:dis}.
\subsection{Representation of the M\"obius group}
Let us now obtain the  representations of the conformal group $SL(2,\C)$ 
on the configuration space of $N$ points on the plane. The
group acts by M\"obius transformation on the space, that is as
\begin{equation}
\label{sl2c}
z\longmapsto z'=\frac{a z+b}{c z+d}, \quad a,b,c,d,z\in\C,\;ad-bc=1,
\end{equation} 
with a similar action on the conjugate variable $\bar{z}$. In two dimensions
the actions on $z$ and $\bar{z}$ may be treated independently. 
We shall display formulas for the holomorphic part only. 

A holomorphic representation of the M\"obius group is 
furnished by the regular functions
on $\C^N$ which transform under $SL(2,\C)$ as
\begin{equation}
\begin{split}
\label{sl2rep}
f(z_1,z_2,\cdots,z_N)
&\longmapsto f(z'_1,z'_2,\cdots,z'_N)\\
&= (c z_1+d)^{\Delta_1}(c z_2+d)^{\Delta_2}\cdots
(c z_N+d)^{\Delta_N} f(z_1,z_2,\cdots,z_N),
\end{split}
\end{equation} 
with ${\boldsymbol\Delta}=(\Delta_1,\Delta_2,\cdots,\Delta_N)$ an $N$-tuple
of real numbers.

Let us first note that the quantities $z_{ij}$ are
equivariant under the M\"obius transformation \eq{sl2c}, 
\begin{equation}
\label{zzp}
z_{ij}\longmapsto z'_{ij}=(c z_i+d)^{-1}(c z_j+d)^{-1} z_{ij}.
\end{equation} 
From \eq{sl2c} we also have
\begin{equation}
	\label{dz:tr}
	dz' = (cz+d)^{-2}dz.
\end{equation}
The integral \eq{IN} is equivariant with respect to \eq{sl2c} 
with degree of homogeneity $-1$ provided $|{\boldsymbol\mu}|=2$. 
In this case it transforms under the M\"obius group as
\begin{equation}
	\label{Itrans}
I_N^{\boldsymbol\mu}(\z)\longmapsto 
I_N^{\boldsymbol\mu}(\z') = 
(c z_1+d)^{\mu_1}(c z_2+d)^{\mu_2}\cdots
(c z_N+d)^{\mu_N} 
I_N^{\m}(\z).
\end{equation} 
Holomorphic representations of the M\"obius group may thus be
constructed out of $z_{ij}$ and $I^{\boldsymbol\mu}_N({\z})$.

We have discussed above the form of $\inmuz$ for $N=2$. 
The expression \eq{I2:gen} with arbitrary parameters does not transform 
under $SL(2,\C)$, while \eq{I2m2} does. 
Equation \eq{Itrans} requires $\mu_1$ and $\mu_2$ to be equal. Thus, from
\eq{I2m2}
\begin{equation}
	\label{I11}
	\inmu{2}{(1,1)}{z_1,z_2} = \frac{\G{0}}{z_{12}}.
\end{equation}
For the other special case $N=3$ equation \eq{laueq} is solved with 
\begin{equation}
	\label{I3}
	\inmu{3}{(\mu_1,\mu_2,\mu_3)}{z_1,z_2,z_3}=
	z_{12}^{-\tfrac{\mu_1+\mu_2-\mu_3}{2}}
	z_{23}^{-\tfrac{\mu_2+\mu_3-\mu_1}{2}}
	z_{31}^{-\tfrac{\mu_3+\mu_1-\mu_2}{2}},
\end{equation}
for $|\m|=\mu_1+\mu_2+\mu_3=2$ up to a multiplicative constant.
This can be verified by plugging the expression into \eq{laueq} and appealing
to the uniqueness of its solution. 

For $N>3$ complications arise due to 
the fact that there exist invariants of the M\"obius transformation, known as
cross ratios, which may be multiplied to any function
with arbitrary exponents without altering the transformation property of
$\inmuz$. This, however, may change the
behavior of functions at infinity on the configuration space. A cross ratio has the form
\begin{equation}
	\label{chi4}
	\chi_{ijkl}=\frac{z_{ij}z_{kl}}{z_{ik}z_{jl}},
\end{equation}
its invariance under M\"obius transformation follows from \eq{zzp}.
It will turn out convenient to denote the cross ratios by
\begin{equation}
	\label{xis}
	\xi_A = \prod_{\substack{i,j\\1\leq i<j\leq N}} z_{ij}^{\alpha^A_{ij}},
\end{equation}
with 
\begin{equation}
	\label{alpha}
	\begin{split}
		\alpha^A_{ji}=\alpha^A_{ij}, \ i<j;\quad
		\sum_{j=1}^N\alpha^A_{ij}=0, \forall i
	\end{split}
\end{equation}
for each $A$.
This will allow treating them rather symmetrically. 
Then, in view of the equivariance \eq{Itrans}, the integral
$\inmuz$ can be written as products of $z_{ij}$ with 
appropriate indices and a function of the cross ratios as
\begin{equation}
	\label{INI0}
	\inmuz = \prod_{\substack{i,j\\1\leq i<j\leq N}} z_{ij}^{\beta_{ij}} I_0(\bxi),
\end{equation}
where $I_0(\bxi)$ is a function of the cross ratios $\bxi=(\xi_1,\xi_2,\cdots)$ and
\begin{equation}
	\label{beta}
	\begin{split}
	\sum_{j=1}^N\beta_{ij}=-\mu_i;
	\quad\beta_{ji}=\beta_{ij},\ i<j,
	\end{split}
\end{equation}
for each $i=1,2,\cdots,N$. Since $|\m|=2$, we also have 
\begin{equation}
\label{betasum}
\sum_{\substack{i,j\\1\leq i<j\leq N}}\beta_{ij}=-1
\end{equation}
Plugging in \eq{INI0} with 
\eq{xis} and \eq{chi4} in \eq{laueq}, we obtain a differential  
equation for the  invariant function $I_0$ of the cross ratios as
\begin{equation}
	\label{eq:I0}
	\begin{split}
		&\sum_{A,B}
		\left(\sum_{\substack{k,l\\1\leq k,l\leq N\\k\neq i, l\neq j}}
		\alpha^A_{ik}\alpha^B_{jl}\chi_{ijkl}\right) 
	\xi_A\xi_B\pa_A\pa_B I_0(\bxi)\\
		&+\sum_A\left(\alpha^A_{ij}
		+\sum_{\substack{k,l\\1\leq k,l\leq N\\k\neq i, l\neq j}}
		\left(\alpha^A_{ik}\alpha^A_{jl}+\alpha^A_{ik}\beta_{jl}
		+\alpha^A_{jl}\beta_{ik}\right)\chi_{ijkl}\right)
	\xi_A\pa_AI_0(\bxi)\\
		&+\left(\beta_{ij}+\sum_{\substack{k,l\\1\leq k,l\leq N\\
			k\neq i, j\neq l}}\beta_{ik}\beta_{jl}\chi_{ijkl}\right)I_0(\bxi)=0,
	\end{split}
\end{equation}
where $\pa_A$ denotes differentiation with respect to $\xi_A$. This equation
is valid for arbitrary $N$.
\subsubsection{Four points}
For four points in two dimensions there is but a single independent cross ratio 
which we choose to be $\xi=\chi_{1234}$. The
non-vanishing exponents $\alpha$ for this choice are
\begin{equation} 
\label{alpha4}
\alpha_{12}
=\alpha_{34}
=-\alpha_{13}
=-\alpha_{24}=1,
\end{equation} 
where we have suppressed the superscript $A$, which is unity in this case. 
Equation \eq{eq:I0} then leads to 
\begin{equation}
	\label{eq:4ptgen}
f_2(\xi)\frac{d^2I_0}{d\xi^2}+f_1(\xi) \frac{dI_0}{d\xi}+f_0(\xi)I_0=0,
\end{equation} 
where
\begin{equation} 
f_2(\xi)=\xi^2(\xi-1)
\end{equation} 
\begin{equation} 
f_1(\xi)=
\xi\big(
(\beta_{13}+\beta_{14} +\beta_{23}+\beta_{24})+
\xi(1-\beta_{13}-\beta_{24})
\big)
\end{equation} 
\begin{equation}
\begin{split}
f_0(\xi)=-\beta_{12}\beta_{34}+\frac{\xi\beta_{14}
\beta_{23}}{\xi-1}+\xi\beta_{13}\beta_{24}
\end{split}
\end{equation} 
This is solved with
\begin{equation}
	\label{I0:solgen}
\begin{split}
I_0(\xi)&=
\xi^{-\beta_{12}}(1-\xi)^{-\beta_{23}}
C_1F(-\beta_{12}-\beta_{13}-\beta_{23},-\beta_{12}-\beta_{23}-\beta_{24},
1-\beta_{12}+\beta_{34};\xi)\\
&\quad+\xi^{-\beta_{34}}(1-\xi)^{-\beta_{23}}
C_2F(1+\beta_{12}+\beta_{13}+\beta_{14},1+\beta_{12}+\beta_{14}+\beta_{24},
1+\beta_{12}-\beta_{34};\xi),
\end{split}
\end{equation} 
where $F$ denotes the Gauss hypergeometric function and
$C_1$ and $C_2$ are arbitrary constants.
The six parameters $\beta$ are related to the weights by the
four equations \eq{beta} through
\begin{equation}
\label{solbeta}
	\begin{split}
		\beta_{12}&=1-\mu_1-\mu_2+\beta_{34},\\
		\beta_{13}&=\mu_2-1-\beta_{14}-\beta_{34},\\
		\beta_{23}&=1-\mu_2-\mu_3+\beta_{14},\\
		\beta_{24}&=-\mu_4-\beta_{14}-\beta_{34}.
	\end{split}
\end{equation}
Plugging in these values along with  \eq{I0:solgen} in \eq{INI0} yields
the four-point integral
\begin{equation}
\label{D2I4}
	\inmu{4}{\m}{z}=z_{12}^{1-\mu_1-\mu_2}z_{13}^{\mu_2-1}
	z_{23}^{1-\mu_2-\mu_3}z_{24}^{-\mu_4}
\big(C_1 F(1-\mu_2,\mu_4;\mu_3+\mu_4;\xi)+
	C_2\xi^{\mu_1+\mu_2-1}
	F(\mu_1,1-\mu_3;\mu_1+\mu_2;\xi)\big).
\end{equation}
with $\xi=z_{12}z_{34}/z_{13}z_{24}$, where we used $|\m|=2$.
\subsubsection{Five points}
Two independent cross ratios exist for five two-dimensional points
which we choose to be 
$\xi_A=\chi_{A,A+1,A+2,A+3}$ for $A=1,2$. 
The non-vanishing exponents are 
\begin{gather} 
\label{alpha5:a}
\alpha^1_{12}=\alpha^1_{34}=-\alpha^1_{13}=-\alpha^1_{24}=1
\\
\label{alpha5:b}
\alpha^2_{23}
=\alpha^2_{45}
=-\alpha^2_{24}
=-\alpha^2_{35}=1.
\end{gather} 
Equation \eq{eq:I0} gives rise to ten equations
for the ten independent choices of the pairs 
$\{(i,j)|i<j;i,j\in(1,2,3,4,5)\}$.
Instead of solving them generally, equation \eq{beta} may be exploited to
set five of the $\beta$'s to zero. We choose 
\begin{equation}
\label{bchus}
\beta_{12}= \beta_{14}= \beta_{15}= \beta_{25}= \beta_{45}=0.
\end{equation} 
The rest are related to the weights by \eq{beta} as
\begin{equation}
\label{beta5mu}
\begin{split}
\beta_{13}&=-\mu_1,\,
\beta_{23}=1-\mu_2-\mu_3,\,
\beta_{24}=\mu_3-1,\\
\beta_{34}&=1-\mu_3-\mu_4,\,
\beta_{35}=-\mu_5.
\end{split}
\end{equation} 
The equations corresponding to the choices $(i,j)=(1,2)$ and
$(i,j)=(4,5)$ ensuing from \eq{eq:I0} are
\begin{equation}
\label{appell:F2}
\begin{split}
\xi_1(1-\xi_1)\frac{\pa^2I_0}{\pa\xi_1^2}
-\xi_1\xi_2\frac{\pa^2I_0}{\pa\xi_1\pa\xi_2}
+\big(c_1-(1+a+b_1)\xi_1\big)
\frac{\pa I_0}{\pa\xi_1}
-b_1\xi_2\frac{\pa I_0}{\pa\xi_2}
-ab_1I_0&=0,\\
\xi_2(1-\xi_2)\frac{\pa^2I_0}{\pa\xi_2^2}
-\xi_1\xi_2\frac{\pa^2I_0}{\pa\xi_1\pa\xi_2}
+\big(c_2-(1+a+b_2)\xi_2\big)
\frac{\pa I_0}{\pa\xi_2}
-b_2\xi_1\frac{\pa I_0}{\pa\xi_1}
-ab_2I_0&=0,
\end{split}
\end{equation} 
where $\xi_1=\tfrac{z_{12}z_{34}}{z_{13}z_{24}}$, 
$\xi_2=\tfrac{z_{23}z_{45}}{z_{24}z_{35}}$ are the cross ratios corresponding
to \eq{alpha5:a} and \eq{alpha5:b}. The parameters are related to
the scaling exponents
\begin{equation}
a=1-\mu_3,\,
b_1=\mu_1,\,
b_2=\mu_5,\,
c_1=\mu_1+\mu_2,\,
c_2=\mu_4+\mu_5,
\end{equation} 
where the sum of the scaling exponents $|\m|=2$.
These are the equations satisfied by
the second Appell hypergeometric function $F_2$. The most general solution,
obtained using \eq{beta5mu} in \eq{INI0} is
\begin{equation}
		\inmu{5}{\m}{\bxi}=z_{13}^{-\mu_1} 
		z_{23}^{1-\mu_2-\mu_3}z_{24}^{\mu_3-1}
	z_{34}^{1-\mu_3-\mu_4}z_{35}^{-\mu_5} I_0(\xi_1,\xi_2),
\end{equation}
where the invariant is
\begin{equation}
\label{I5}
	\begin{split}
		I_0(\xi_1,\xi_2)&=
		C_1 F_2(1-\mu_3,\mu_1,\mu_5,\mu_1+\mu_2,\mu_4+\mu_5;\xi_1,\xi_2)\\
		&\qquad +C_2 \xi_1^{1-\mu_1-\mu_2} 
		F_2(\mu_4+\mu_5,1-\mu_2,\mu_5,\mu_3+\mu_4+\mu_5,\mu_4+\mu_5;
		\xi_1,\xi_2)\\
		&\qquad +C_3 \xi_2^{1-\mu_4-\mu_5} 
		F_2(\mu_1+\mu_2,\mu_1,1-\mu_4,\mu_1+\mu_2,\mu_1+\mu_2+\mu_3;
		\xi_1,\xi_2)\\
		&\qquad +C_4 \xi_1^{1-\mu_1-\mu_2}\xi_2^{1-\mu_4-\mu_5} 
		F_2(1,1-\mu_2,1-\mu_4,\mu_3+\mu_4+\mu_5,\mu_1+\mu_2+\mu_3;
		\xi_1,\xi_2),
	\end{split}
\end{equation} 
where $C_1$, $C_2$, $C_3$ and $C_4$ are arbitrary constants.
As in the case of four points, the final result does not depend on the
choice of $\beta$'s in \eq{bchus}. The other eight equations obtained from
\eq{eq:I0} pairwise yield the equations for the Appell function $F_2$ in
other domains, related to the present one by analytic continuation. 

Above considerations as well as all the expressions have anti-holomorphic
counterparts with $\mu$ changed to $\mu'$.
\subsection{Correlation functions}
Correlation functions in two-dimensional conformal field theories are
well-known. We repeat some of the computations here in order to 
bring out the analogy with the four-dimensional
counterpart. For this purpose it suffices to consider chiral primary 
scalar fields $\{\phi_i(z_i)\}$  
with conformal dimensions $\boldsymbol\Delta$ in line with \eq{sl2rep}.
The correlation function of $N$ chiral scalar primaries is given
by a holomorphic
representation \eq{sl2rep} on the configuration space $\conf_N(\C)$. 
In particular, it is invariant under
translation. The anti-holomorphic part follows suit  with conjugated 
coordinates and primed weights. It then follows from
the preceding discussion that a correlation function 
for chiral primaries can be expressed in terms of the differences
$z_{ij}$ and the integrals $\inmu{N}{\m}{\z}$. Since $SL(2,\C)$ equivariance restricts the degree
of homogeneity of the integrals to be $-1$ by constraining $|\m|=2$, we can write down correlation
functions of a set of primary fields with given conformal dimensions by simply multiplying the 
integrals by powers of $z_{ij}$ so as to satisfy \eq{sl2rep},
\begin{equation}
\label{gn}
G^{\Delta_1,\Delta_2,\cdots,\Delta_N}_N(\phi_1,\phi_2,\cdots\phi_N) =
\mathcal{F}\left(
\prod\limits_{\substack{i,j\\1\leq i<j\leq N}}
z_{ij}^{\ell_{ij}}
	I_N^{\boldsymbol\m}({\boldsymbol{z}})\right),
\end{equation} 
where $\mathcal{F}$ indicates a functional involving sums and integrals
of $I$ with respect to its parameters, transforming appropriately under the 
M\"obius group. We use the shorthand $\phi_i$ for $\phi(z_i)$. The parameters
$\ell$ are related to the weights and conformal dimensions of fields as
\begin{equation}
\label{amd}
-\!\!\sum\limits_{\stackrel{j}{1\leq j\leq N}; j\neq i}
\ell_{ij}+\mu_i=\Delta_i,\quad
\sum\limits_{i=1}^N\mu_i=2
\end{equation} 
for each $i=1,2,\cdots,N$ and we have defined $\ell_{ji}=\ell_{ij}$ if
$j>i$. The product  in front of the integral in \eq{gn} 
is referred to as the leg factor. We shall suppress the superscripts in $G_N$
if the conformal dimensions involved are clear from the context.

The correlation functions for $N=2$ and $N=3$, the two-point and three-point
functions, respectively, are fixed up to a constant by their $SL(2,\C)$
equivariance.
For example, by \eq{gn},
\begin{equation}
	G^{\Delta_1,\Delta_2}_2(\phi_1,\phi_2)=z_{12}^{\ell_{12}}\inmu{2}{(1,1)}{z_1,z_2},
\end{equation} 
and we have, by \eq{amd},
\begin{equation}
\ell_{12}=1-\Delta_1=1-\Delta_2,
\end{equation} 
It follows, in accordance with \eq{sl2rep}, 
that $\Delta_1=\Delta_2$. Using \eq{I11} 
we thus obtain
\begin{equation}
\label{G2pt}
	G^{\Delta_1,\Delta_2}_2(\phi_1,\phi_2)=\frac{C_{\Delta_1}
\G{0}}{z_{12}^{\Delta_1}}\delta_{\Delta_1,\Delta_2},
\end{equation} 
where $C_{\Delta_1}$ is an arbitrary constant for each field of conformal
dimension $\Delta$.
Similarly, for the three-point function
\begin{equation}
G^{\Delta_1,\Delta_2,\Delta_3}_3(\phi_1,\phi_2,\phi_3) 
= C_{\Delta_1,\Delta_2,\Delta_3}
z_{12}^{\ell_{12}}
z_{13}^{\ell_{13}}
z_{23}^{\ell_{23}} 
I_3^{(\mu_1,\mu_2,\mu_3)}(z_1,z_2,z_3).
\end{equation} 
Then by \eq{amd} the exponents of the leg factor satisfy the three equations
\begin{equation}
\begin{split}
\ell_{12}+\ell_{13}=\mu_1-\Delta_1,\\
\ell_{12}+\ell_{23}=\mu_2-\Delta_2, \\
\ell_{13}+\ell_{23}=\mu_3-\Delta_3,
\end{split} 
\end{equation} 
which are solved to obtain
\begin{equation}
\begin{split}
\ell_{12}=
-\frac{1}{2}(\Delta_1+\Delta_2-\Delta_3)+ \frac{1}{2}(\mu_1+\mu_2-\mu_3)\\
\ell_{13}=
-\frac{1}{2}(\Delta_1+\Delta_3-\Delta_2)+ \frac{1}{2}(\mu_1+\mu_3-\mu_2)\\
\ell_{23}=
-\frac{1}{2}(\Delta_2+\Delta_3-\Delta_1)+ \frac{1}{2}(\mu_2+\mu_3-\mu_1).
\end{split}
\end{equation} 
Using \eq{I3} for $I_3$ then yields the three-point function
\begin{equation}
	\label{G3}
G^{\Delta_1,\Delta_2,\Delta_3}_3(\phi_1,\phi_2,\phi_3)
=C_{\Delta_1\Delta_2\Delta_3}
z_{12}^{-\frac{1}{2}(\Delta_1+\Delta_2-\Delta_3)}
z_{13}^{-\frac{1}{2}(\Delta_1+\Delta_3-\Delta_2)}
z_{23}^{-\frac{1}{2}(\Delta_2+\Delta_3-\Delta_1)}.
\end{equation} 
Let us remark that in these two cases the integrals did not have a role 
to play. The leg factor in
both cases were so arranged as to obviate the $\mu$'s, thereby
effacing the trace of the integrals.
Thus, the two- and three-point correlation function of primaries are completely determined by
their equivariance under the M\"obius group and the given conformal dimensions. This does not 
generalize to higher point functions, however. While the leg factors could be so arranged as to
annul the contributions of $\beta$'s in \eq{INI0}, the cross ratios introduce arbitrariness in
the leg factors. This calls for further restrictions on the correlation functions. One such 
stipulation arises from requiring that higher point functions can be pared
down to products of three-point functions, which we now proceed to discuss.
\subsection{Projectors}
Parsing of higher point 
correlation function in terms of the three-point function 
is effected by using projectors \cite{fgpg,sd,do}. 
There is an appropriate set of projectors $\{\Pid\}$ 
summing up to the identity operator $\I$ 
\begin{equation}
	\label{proj}
	\I=\sum_{\Delta}\Pid,
\end{equation}
such that, the $N$-point function can be parsed as
\begin{equation}
	\begin{split}
		G_N(\phi_1,\phi_2,\cdots,\phi_N) &=\langle\phi_1\phi_2\cdots\phi_N\rangle\\
		&=\langle\phi_1\phi_2\I\phi_3\I\phi_4\cdots\phi_{N-2}\I\phi_{N-1}\phi_N\rangle\\
		&=\sum_{\Delta,\Delta',\cdots,\Delta''}
		\langle\phi_1\phi_2\Pi_{\Delta}
		\phi_3\Pi_{\Delta'}\phi_4\cdots\phi_{N-2}\Pi_{\Delta''}
		\phi_{N-1}\phi_N\rangle,
	\end{split}
\end{equation}
where $\phi_{\Delta}(x)$ denotes a primary field of conformal dimension $\Delta$ at $x\in\C$.
The operator $\Pid$ is defined as
\begin{equation}
	\label{PidN}
	\Pid=
	\frac{1}{N_{\Delta}}\int\frac{\phi_{\Delta}(x)
	\phi_{\Delta}(y)}{(x-y)^{2-\Delta}} dx dy,
\end{equation}
where $N_{\Delta}$ is a constant and the path of integration, 
written formally in this 
expression, is fixed only when used in conjunction with a correlation function.
By \eq{zzp} and \eq{dz:tr}, $\Pi_{\Delta}$ is invariant under the M\"obius
group.
The constant is determined by requiring the projector to be consistent 
with two point functions. The composition of the projectors is defined as
\begin{equation}
	\Pid\circ\Pi_{\Delta'} = \frac{1}{N_{\Delta}N_{\Delta'}} 
	\int\frac{\phi_{\Delta}(x) G_2\left(\phi_{\Delta}(y)\phi_{\Delta'}(x')\right)
	\phi_{\Delta'}(y')}{(x-y)^{2-\Delta}(x'-y')^{2-\Delta'}}
	dxdydx'dy',
\end{equation}
which also defines their action on fields in parsing the correlation
function. Using \eq{G2pt} this yields
\begin{equation}
	\Pid\circ\Pi_{\Delta'} = {\delta_{\Delta,\Delta'}}\frac{C_{\Delta}
          \G{0}}{N_{\Delta}^2} 
	\int\frac{\phi_{\Delta}(x) \phi_{\Delta}(y')}{%
		(x-y)^{2-\Delta}(x'-y')^{2-\Delta}(y-x')^{\Delta}}
	dxdydx'dy'.
\end{equation}
The integral over $x'$ can be performed using \eq{I2m2}. This leads to 
\begin{equation}
	\Pid\circ\Pi_{\Delta'} = {\delta_{\Delta,\Delta'}}
	\frac{C_{\Delta}\G{1-\Delta}\G{\Delta-1}}{N_{\Delta}^2} 
	\int{\phi_{\Delta}(x) \phi_{\Delta}(y')}
	dxdy'
	\inmu{2}{(2-\Delta,1)}{x,y'}
\end{equation}
Let us note that the integral $\inmu{2}{(2-\Delta,1)}{x,y'}$ 
appearing in this expression does
not have $|\m|=2$. Hence it is not well-behaved at infinity. 
The final result is indeed conformal invariant.
Using \eq{I2:gen}, we evaluate the above product to be 
\begin{equation}
	\Pid\circ\Pi_{\Delta'} = {\delta_{\Delta,\Delta'}}
	\frac{C_{\Delta}\G{0}\G{1-\Delta}\G{\Delta-1}}{%
N_{\Delta}^2} 
	\int\frac{\phi_{\Delta}(x) \phi_{\Delta}(y')}{(x-y')^{2-\Delta}}
	dxdy'
\end{equation}
Since a projector is idempotent, equating to \eq{PidN} we obtain
\begin{equation}
	N_{\Delta}={C_{\Delta}
\G{0}\G{1-\Delta}\G{\Delta-1}}.
\end{equation}
We have absorbed factors of powers of $(-1)$ in the constant $C_{\Delta}$.
The apparent lack of convergence of the projector is due to the 
unspecified nature of the sum over $\Delta$ in \eq{proj}.
\subsection{Higher points correlation functions}
Let us now use the projectors to express higher point correlation functions in terms of the Lauricella 
functions. We shall demonstrate this for $N=4$ and $N=5$.

The four point correlation function $G_4(z_1,z_2,z_3,z_4)=\langle\phi_1\phi_2\phi_3\phi_4\rangle$ is 
written by inserting the projector \eq{proj} with \eq{PidN} as
\begin{equation}
	\begin{split}
		G_4(\phi_1,\phi_2,\phi_3,\phi_4)&=
		\sum_{\Delta}\frac{1}{N_{\Delta}} \int\frac{dxdy}{(x-y)^{2-\Delta}}
		\langle\phi_1\phi_2\phi_{\Delta}(x)\rangle
	\langle\phi_{\Delta}(y)\phi_3\phi_4\rangle\\ 
		&=\sum_{\Delta}\frac{1}{N_{\Delta}} \int\frac{dxdy}{(x-y)^{2-\Delta}}
G_3^{\Delta_1,\Delta_2,\Delta}(z_1,z_2,x) 
G_3^{\Delta,\Delta_3,\Delta_4}(y,z_3,z_4) .
	\end{split}
\end{equation}
We have thus expressed the four point function in terms of the three 
point functions. Expanding the latter using
\eq{G3} we first collect all the terms containing the integration variable $x$. They combine into 
$\inmu{3}{\left(2-\Delta,\tfrac{\Delta+\Delta_1-\Delta_2}{2},\tfrac{\Delta_2
+\Delta-\Delta_1}{2}\right)}{y,z_1,z_2}$, whose weights add up to $2$. 
Using \eq{I3} this furnishes powers of two linear forms in $y$, which 
combined with the two more from the second $G_3$ factor in the last integral 
leaves us with  an integral in $y$ with a total of 
four factors of powers of linear forms in $y$ in the integrand. 
Collecting all yields
\begin{equation}
	\label{d2G4}
	\begin{split}
		G_4(\phi_1,\phi_2,\phi_3,\phi_4)&=
		\sum_{\Delta}
              \frac{C_{\Delta_1\Delta_2\Delta}C_{\Delta,\Delta_3,\Delta_4}}{%
              N_{\Delta}} 
		z_{12}^{-\tfrac{\Delta_1+\Delta_2+\Delta-2}{2}}
		z_{34}^{-\tfrac{\Delta_3+\Delta_4-\Delta}{2}}\\
		&\qquad\times\inmu{4}{\left(
			\tfrac{2-\Delta+\Delta_1-\Delta_2}{2},
			\tfrac{2-\Delta+\Delta_2-\Delta_1}{2},
			\tfrac{\Delta+\Delta_3-\Delta_4}{2},
			\tfrac{\Delta+\Delta_4-\Delta_3}{2}\right)
			}{z_1,z_2,z_3,z_4}.
	\end{split}
\end{equation}
The five point function is similarly parsed as
\begin{equation}
	\begin{split}
		G_5(\phi_1,\phi_2,\phi_3,\phi_4,\phi_5)&=
		\sum_{\Delta,\Delta'}\frac{1}{N_{\Delta}N_{\Delta'}} \int
		\frac{\langle\phi_1\phi_2\phi_{\Delta}(x)\rangle
		\langle\phi_{\Delta}(y)\phi_3\phi_{\Delta'}(x')\rangle 
		\langle\phi_{\Delta'}(y')\phi_4\phi_5\rangle}
			{(x-y)^{2-\Delta}(x'-y')^{2-\Delta'}} dxdydx'dy'\\
		&=\sum_{\Delta,\Delta'}\frac{1}{N_{\Delta}N_{\Delta'}} 
		\int\frac{
		G_3^{\Delta_1,\Delta_2,\Delta}(z_1,z_2,x) 
		G_3^{\Delta,\Delta_3,\Delta'}(y,z_3,x') 
		G_3^{\Delta',\Delta_4,\Delta_5}(y',z_4,z_5)} 
			{(x-y)^{2-\Delta}(x'-y')^{2-\Delta'}} dxdydx'dy'.
	\end{split}
\end{equation}
Using \eq{I3} repeatedly and performing integrals in turn until the 
integration over only a single 
variable is left, this is finally written in terms of $I_5$ as
\begin{equation}
\label{G5}
\begin{split}
		G_5(\phi_1,\phi_2,\phi_3,\phi_4,\phi_5)&=
\left(\frac{1}{2\pi i}\right)^3
		\sum_{\Delta,\Delta'}
\frac{C_{\Delta_1\Delta_2\Delta}
C_{\Delta\Delta_3\Delta'}C_{\Delta'\Delta_4\Delta_5}}{%
N_{\Delta}N_{\Delta'}} 
z_{12}^{-\tfrac{1}{2}(\Delta_1+\Delta_2-\Delta)}
z_{45}^{-\tfrac{1}{2}(\Delta_4+\Delta_5+\Delta'-2)}\\
&\times\int ds_1ds_2ds_3
\frac{\G{-s_1}\G{-s_2}\G{-s_3}\G{m+s_1}\G{n+s_2}\G{p+s_3}}{%
\G{m}\G{n}\G{p}}\\
&\times\int d\tau
(-\tau)^{-(s_1+s_2+s_3+2)}\\&
\times I_5^{%
\big(\tfrac{1}{2}(\Delta_1-\Delta_2+\Delta),
\tfrac{1}{2}(\Delta_2-\Delta_1+\Delta),
\tfrac{1}{2}(2+\Delta_3-\Delta-\Delta')-s_1,
-s_2,-s_3
\big)}%
(z_1,z_2,z_3,z_4,z_5),
\end{split}
\end{equation} 
where the quantities
\begin{equation} 
m=\frac{1}{2}(\Delta_3+\Delta+\Delta'-2),\quad
n=\frac{1}{2}(\Delta_4-\Delta_5+2-\Delta'),\quad
p=\frac{1}{2}(\Delta_5-\Delta_4+2-\Delta')
\end{equation} 
have been defined and repeated use of the integral
\begin{equation}
\frac{1}{(1-x)^n}=\frac{1}{2\pi i}\frac{1}{\G{n}}\int_{-i\infty}^{i\infty}
ds (-x)^s\G{-s}\G{n+s}
\end{equation} 
has been made.

Correlation functions with more number of 
points can be similarly written down in terms
of the Lauricella functions $I_N$. 
We have thus related the conformal correlation functions of scalar primaries
to the Lauricella system, defined on the configuration space of points in two 
dimensions.
\section{Four dimensions}
The conformal or M\"obius group of the compactified four-dimensional 
Euclidean space 
$M=\R^4\cup\{\infty\}$ is $SL(2,\H)$ \cite{porteous,baez}. 
The correlation functions of scalar primaries of a four-dimensional conformal
field theory are obtained as representations of $SL(2,\H)$ on the
configuration space of $N$ points in $M$.
In this section we show that the 
considerations of the previous section carry over 
\emph{mutatis mutandis} to the 
four-dimensional Euclidean conformal field theories. 
In order to fix notations let us begin by recalling some facts about 
quaternions and the M\"obius transformations \cite{wilker,dobrev,petkova}. 
\subsection{Quaternions}
A quaternion can be written as a $2\times 2$ matrix with a pair  of complex numbers $U$ and $V$
and their respective complex conjugates $\bar{U}$ and $\bar{V}$ as
\begin{equation}
	\label{qmat}
	Q=\begin{pmatrix}
		U&V\\-\bar{V}&\bar{U}
	\end{pmatrix}\in\H,\quad U,V\in\C.
\end{equation}
The norm squared of a quaternion is 
\begin{equation}
	\norm{Q}^2 = Q\Qd=|Q|=U\bar{U}+V\bar{V}, 
\end{equation}
where $\Qd$ denotes the Hermitian conjugate and 
$|Q|$ denotes the determinant of the matrix \eq{qmat}.
The inverse of the matrix $Q$ is
\begin{equation}
	\label{qinv}
	Q^{-1} = \frac{1}{|Q|}Q^{\dagger}.
\end{equation}
A quaternion can also be looked upon as a Euclidean real
four-vector  $\q=(q_0,q_1,q_2,q_3)$ by writing $U=q_0+iq_3$ and 
$V=q_1+iq_2$. From \eq{qmat}, then, 
\begin{equation}
	\label{Q2}
	Q=\begin{pmatrix}
		q_0+iq_3&q_1+iq_2\\-q_1+iq_2&q_0-iq_3
	\end{pmatrix}
\end{equation}
The norm-squared of the quaternion $Q$ is the Euclidean norm-squared 
of the four-vector, 
\begin{gather}
	\label{norm}
	\norm{Q}^2 = |Q|= q^2 
	= \q\cdot\q = q_0^2+q_1^2+q_2^2+q_3^3.
\end{gather}
The volume form of the four-dimensional Euclidean space is then written as
the wedge product of the column vectors of the 
differential of $Q$ divided by $2^4$,
\begin{equation}
		\label{vol4}
	\begin{split}
		d^4Q &=d^4q=dq_0\wedge dq_1\wedge dq_2\wedge dq_3.
	\end{split}
\end{equation}
This generalizes the two-dimensional volume form $dz\wedge d\bar{z}$. 
In the previous section we
chose to only write the holomorphic parts to leave provision for spin. 
In four dimensions we need
to consider four-dimensional integrals. We consider integrals 
similar to \eq{IN} in 
four dimensions. We shall denote these by the same symbol as in \eq{IN}. 
Let us define
\begin{equation}
	\label{INQ}
	\inmu{N}{\m}{\Q} = \int\frac{d^4Q}{
		|Q-Q_1|^{\mu_1}|Q-Q_2|^{\mu_2}\cdots|Q-Q_N|^{\mu_N}},
\end{equation}
where $\Q$ denotes the $N$-tuple of quaternions, $\Q=(Q_1,Q_2,\cdots,Q_N)$. From \eq{qinv} we have
\begin{equation}
	\frac{\pa|Q|}{\pa Q}=\Qd=|Q|Q^{-1}.
\end{equation}
For the following it is useful to indicate the matrix indices of the 
quaternions, $Q=(Q)_{ab}$ and $Q^{-1}=(Q)^{ba}$,
$1\leq a,b\leq 2$. Then $(Q)_{ab}(Q)^{bc}=\delta^c_a$ and the last equation
becomes
\begin{equation}
	\frac{\pa|Q|}{\pa (Q)_{ab}}=|Q|(Q)^{ba}.
\end{equation}
Using this expression for the derivative of the determinant and the identity
\begin{equation}
	\label{identityQ}
	(Q-Q_i)^{-1}(Q_i-Q_j)(Q-Q_j)^{-1}=(Q-Q_i)^{-1}-(Q-Q_j)^{-1},
\end{equation}
which generalizes \eq{identity3}, we obtain, by differentiating under 
the integral sign in \eq{INQ} a differential equation
\begin{equation}
	\label{lau:Q}
	\sum_{b,c=1}^2(Q_{ij})_{bc}\frac{\pa}{\pa (Q_i)_{ba}}
	\frac{\pa\inmu{N}{\m}{\Q}}{\pa (Q_j)_{dc}} =
	\mu_j\frac{\pa\inmu{N}{\m}{\Q}}{\pa (Q_i)_{da}}
	-\mu_i\frac{\pa\inmu{N}{\m}{\Q}}{\pa (Q_j)_{da}},
\end{equation}
where $i,j=1,2,\cdots,N$
and we used the abbreviation $Q_{ij}=Q_i-Q_j$. 
This equation generalizes \eq{laueq}. We refer to this as the quaternionic 
Lauricella system. 
Let us stress that the order of quaternions are important in these formulas, 
since they are  non-commutative and represented here as complex matrices. 

As in two dimensions, $N=2$ and $N=3$ are special. 
Let us discuss them first. We have, using \eq{norm} in \eq{INQ}
\begin{equation}
	\inmu{2}{(\mu_1,\mu_2)}{Q_1,Q_2} = \int \frac{d^4q}{%
		(q-q_1)^{2\mu_1}(q-q_2)^{2\mu_2}},
\end{equation}
which is evaluated using Feynman parametrization of the integrand to be
\begin{equation}
	\label{IQ2mm}
	\inmu{2}{(\mu_1,\mu_2)}{Q_1,Q_2} = 
	\frac{\pi^2\G{{2}-\mu_1}\G{{2}-\mu_2}
	\G{\mu_1+\mu_2-{2}}}{\G{\mu_1}\G{\mu_2}\G{4-\mu_1-\mu_2}}
	\frac{1}{\left|Q_{12}\right|^{\mu_1+\mu_2-2}},
\end{equation}
It can be verified that this 
satisfies \eq{lau:Q}. 
Let us note that it is translation invariant and homogeneous with degree
$2-|\m|$.
This expression generalizes \eq{I2:gen} with doubled
numbers reflecting the doubling of dimension from two to four.

\subsection{Representation of the M\"obius group}
\label{mob}
The conformal group of $\R^4\cup\{\infty\}$ is isomorphic to the group of 
$2\times 2$ matrices whose blocks are quaternions, namely, $SL(2,\H)$
\cite{wilker,porteous}. 
We have,
\begin{equation}
	\label{sl2h:def}
	SL(2,\H) = \left\{\left.\begin{pmatrix}A&B\\C&D\end{pmatrix}
		\right| |AC^{-1}DC-BC|=1; A,B,C,D\in\H\right\}.
\end{equation}
The matrix whose determinant is set to unity in this definition can be 
written in seven alternative forms \cite{wilker}. We shall have 
occasion to use only the present one.  
The M\"obius group acts on a quaternion $Q$ similarly as the fractional linear 
transformation \eq{sl2c}, 
\begin{equation}
	\label{sl2h}
	Q'=(AQ+B)(CQ+D)^{-1}.
\end{equation}
Representation of the M\"obius group is furnished by complex-valued
functions of quaternions transforming as,
\begin{equation}
\begin{split}
\label{sl2hrep}
f(Q_1,Q_2,\cdots,Q_N)
&\longmapsto f(Q'_1,Q'_2,\cdots,Q'_N)\\
&= |C Q_1+D|^{\Delta_1}|C Q_2+D|^{\Delta_2}\cdots
|C Q_N+D|^{\Delta_N} f(Q_1,Q_2,\cdots,Q_N),
\end{split}
\end{equation} 
where ${\boldsymbol\Delta}=(\Delta_1,\Delta_2,\cdots,\Delta_N)$ denotes
the the $N$-tuple of weights, as before. 
Generalizing the transformation \eq{zzp} of $z_{ij}$, 
the difference of two quaternions transform under the M\"obius group as
\begin{equation}
	\label{qij:transf}
	\begin{split}
		Q'_{ij}&=(AQ_i+B)(CQ_i+D)^{-1}-(AQ_j+B)(CQ_j+D)^{-1}\\
		&=
		\left((AQ_i+B)-AC^{-1}(CQ_i+D)\right)(CQ_i+D)^{-1}\\
		&\qquad\qquad
		-\left((AQ_j+B)-AC^{-1}(CQ_j+D)\right)(CQ_j+D)^{-1}\\
		&=(AC^{-1}D-B)(CQ_j+D)^{-1}CQ_{ij}(CQ_i+D)^{-1},
	\end{split}
\end{equation}
where we used the identity \eq{identityQ} in the last step.
Taking the determinant of the matrices on both sides and using the fact that the
determinant in \eq{sl2h:def} is unity, we obtain \cite{wilker}
\begin{equation}
	\label{QQp}
	|Q'_{ij}| = |CQ_i+D|^{-1}|CQ_j+D|^{-1}|Q_{ij}|.
\end{equation}
Let us derive the transformation of the volume element, 
generalizing \eq{dz:tr}.
The differential of $Q'$, obtained from \eq{sl2h} is
\begin{equation}
	\begin{split}
		dQ' &= AdQ (CQ+D)^{-1}+(AQ+B)d(CQ+D)^{-1}
	\end{split}
\end{equation}
Since $dM^{-1}=-M^{-1}dMM^{-1}$ for any matrix $M$, we obtain
\begin{equation}
	\label{dQQ}
	\begin{split}
		dQ' &= \left(A-(AQ+B)(CQ+D)^{-1}C\right)dQ(CQ+D)^{-1}\\
		&=\left(AC^{-1}(CQ+D)-(AQ+B)\right)(CQ+D)^{-1}C dQ(CQ+D)^{-1}\\
		&=(AC^{-1}D-B)(CQ+D)^{-1}CdQ(CQ+D)^{-1}\\
		&=(AC^{-1}DC-BC)C^{-1}(CQ+D)^{-1}CdQ(CQ+D)^{-1}.
	\end{split}
\end{equation}
We have thus a relation between the quaternion differentials as
\begin{equation}
	\label{dQprim}
dQ'=XdQY, \quad X,Y\in\H,
\end{equation}
where the quaternions are expressed as $2\times 2$ matrices.
In order to obtain the transformation of the volume form \eq{vol4} it is 
convenient to go over to the four-vector $\q$, written as a column matrix.
A transformation of a 
quaternion by another $dQ\longmapsto XdQ$ given in the $2\times 2$ form can be 
written as a transformation of a four-vector as
\begin{equation}
	\begin{pmatrix}
		dq_0\\dq_1\\dq_2\\dq_3
	\end{pmatrix}
	\longmapsto 
	\begin{pmatrix}
		x_0&-x_1&-x_2&-x_3\\x_3&-x_2&x_1&x_0\\
		x_1&x_0&-x_3&x_2\\x_2&x_3&x_0&-x_1
	\end{pmatrix}
	\begin{pmatrix}
		dq_0\\dq_1\\dq_2\\dq_3
	\end{pmatrix}.
\end{equation}
The determinant of the $4\times 4$ transformation 
matrix equals $|X|^2$. 
The volume form \eq{vol4} obtained by taking wedge product of the components,
transforms under this as
\begin{equation}
	d^4q'=|X|^2d^4q.
\end{equation}

Similarly, a transformation of a 
quaternion by another $dQ\longmapsto dQY$ from the right,
given in the $2\times 2$ form can be 
written as a transformation of the four-vector as
\begin{equation}
	\begin{pmatrix}
		dq_0\\dq_1\\dq_2\\dq_3
	\end{pmatrix}
	\longmapsto 
	\begin{pmatrix}
		y_0&-y_1&-y_2&-y_3\\y_3&y_2&-y_1&y_0\\
		y_1&y_0&y_3&-y_2\\y_2&-y_3&y_0&y_1
	\end{pmatrix}
	\begin{pmatrix}
		dq_0\\dq_1\\dq_2\\dq_3
	\end{pmatrix}.
\end{equation}
The determinant of the $4\times 4$ transformation matrix equals $|Y|^2$. 
Hence the volume form \eq{vol4} transforms as
\begin{equation}
	d^4q'=|Y|^2d^4q.
\end{equation}
Thus, under a transformation \eq{dQprim} the volume form transforms as
\begin{equation}
	\label{dQ:tr}
	d^4q'=|X|^2|Y|^2d^4q.
\end{equation}
Using this for the transformation \eq{dQQ} along with the unity of the 
determinant of the first factor as in the definition \eq{sl2h:def}, we obtain
\begin{equation}
	\label{dQ:tr}
	d^4q'=|CQ+D|^{-4}d^4q.
\end{equation}
The exponent $4$ is the dimension of the space, as did was $2$ in 
\eq{dz:tr}.
Using \eq{QQp} and \eq{dQ:tr} we conclude that the
integral \eq{INQ} is equivariant with degree of homogeneity $-2$, equal
to the dimension of the space, provided 
$|{\boldsymbol\mu}|=4$, as can be verified by transforming 
the $Q_i$ as well as the variable of integration $Q$ in \eq{INQ} according to
\eq{sl2h}, yielding
\begin{equation}
	\label{IQtrans}
I_N^{\boldsymbol\mu}(\Q') = 
|CQ_1+D|^{\mu_1}|C Q_2+D|^{\mu_2}\cdots
|CQ_N+D|^{\mu_N} 
I_N^{\m}(\Q),
\end{equation}
with $|\m|=4$. 
Representations of the M\"obius group $SL(2,\H)$ may thus be
constructed out of $|Q_{ij}|$ and $I^{\boldsymbol\mu}_N({\Q})$.

As in the two-dimensional case,  \eq{IQtrans} requires equality of 
$\mu_1$ and $\mu_2$ for $N=2$, along with $\mu_1+\mu_2=4$, to 
be equivariant. Thus, from
\eq{IQ2mm} we derive the equivariant expression
\begin{equation}
	\label{IQ11}
	\inmu{2}{(1,1)}{Q_1,Q_2} = \frac{\pi^2\G{0}}{|Q_{12}|^2}.
\end{equation}
For the other special case $N=3$, the equation \eq{lau:Q} is solved with 
\begin{equation}
	\label{IQ3}
	\inmu{3}{(\mu_1,\mu_2,\mu_3)}{z_1,z_2,z_3}=
	|Q_{12}|^{-\tfrac{\mu_1+\mu_2-\mu_3}{2}}
	|Q_{23}|^{-\tfrac{\mu_2+\mu_3-\mu_1}{2}}
	|Q_{31}|^{-\tfrac{\mu_3+\mu_1-\mu_2}{2}},
\end{equation}
up to a multiplicative constant and $\mu_1+\mu_2+\mu_3=4$. 
As in the two-dimensional case,
this can be verified by plugging the solution into \eq{lau:Q}.

For $N>3$ complications as in two dimensions arise due to the
existence of cross-ratios.
These are invariants of the $SL(2,\H)$ transformation. 
Considering a product of the determinants of the quaternions
$\prod_{i,j=1}^N |Q_{ij}|$, we recall that it transforms according to
\eq{QQp}. Writing a matrix with entries showing the order of transformation
of $Q_{ij}$ in $Q_i$ along the rows, the
invariants are given by the vectors in its kernel. For example, for $N=4$ 
the matrix of exponents is 
\begin{equation}
	\label{alphamat}
\mathcal{M}=\bordermatrix{%
& Q_{12}& Q_{13} & Q_{14} & Q_{23} & Q_{24} & Q_{34}\cr
Q_1& 1&1&1&0&0&0\cr
Q_2& 1&0&0&1&1&0\cr
Q_3& 0&1&0&1&0&1\cr
Q_4& 0&0&1&0&1&1\cr
}.%
\end{equation} 
The kernel of this matrix is two-dimensional. We can choose the 
basis vectors of the kernel as the transpose of 
\begin{equation}
	\label{2inv}
\bordermatrix{%
& Q_{12}& Q_{13} & Q_{14} & Q_{23} & Q_{24} & Q_{34}\cr
& 1&-1&0&0&-1&1\cr
&0&-1&1&1&-1&0
},%
\end{equation} 
where we indicated the quaternions.
Two invariants are correspondingly given by 
$\tfrac{|Q_{12}||Q_{34}|}{|Q_{13}||Q_{24}|}$ and 
$\tfrac{|Q_{14}||Q_{23}|}{|Q_{24}||Q_{13}|}$.
Generally, for $N$ quaternions, the matrix of exponents is
$N\times\tfrac{N(N-1)}{2}$. Its kernel has dimension $\tfrac{N(N-3)}{2}$,
which is the number of independent invariants that can be constructed from the
determinant of the quaternions. 
The counting in two dimensions was similar, but the Pl\"ucker
relations among the invariants further reduced their number. Thus, for
$N=4$ there was but a single invariant, as we dealt with before, but in four
dimensions there are two invariants for $N=4$. 
Let us first define another set of  quaternions
\begin{equation}
	\label{chi4Q}
	\chi_{ijkl}=Q_{ij}Q_{ik}^{-1}Q_{kl}Q_{jl}^{-1}.
\end{equation}
The determinants of these matrices 
are invariant under $SL(2,\H)$ thanks to \eq{QQp}.
Determinants of all the $\chi$'s are, however, not independent. 
A choice for the 
independent ones is to be made, thereby fixing the asymptotic 
behavior of the integrals. These are taken to be
the cross ratios, the rest being 
functions of them. 
We shall denote the cross ratios as before
\begin{equation}
	\label{xiQ}
	\xi_A = \prod_{\stackrel{i,j}{1\leq i<j\leq N}}
	|Q_{ij}|^{\alpha^A_{ij}},
\end{equation}
where $\alpha^A$ for each $A$ designates a basis 
vector in the kernel of the matrix $\mathcal{M}$, as the rows in \eq{2inv}, for
example. These satisfy \eq{alpha} as before.

Let us denote the trace of the $2\times 2$ matrices $\chi$ by
\begin{equation}
	\label{trchi}
	\tau_{ijkl}=\tr\chi_{ijkl}.
\end{equation}
Then, in view of the equivariance \eq{IQtrans} of $\inmu{N}{\m}{\Q}$, 
it can be written as products of $|Q_{ij}|$ with 
appropriate indices and a function of the cross ratios as
\begin{equation}
	\label{INQ0}
	\inmu{N}{\m}{\Q} = \prod_{\substack{i,j\\1\leq i<j\leq N}} 
	|Q_{ij}|^{\beta_{ij}} I_0(\bxi),
\end{equation}
where $I_0(\bxi)$ is a function of the $N(N-3)/2$ cross ratios 
$\bxi=(\xi_1,\xi_2,\cdots,\xi_{\tfrac{N(N-3)}{2}})$  
and  the $\beta$'s satisfy \eq{beta}, while
\eq{betasum} is replaced with 
\begin{equation} 
\label{betasumQ}
\sum_{\substack{i,j\\1\leq i<j\leq N}}\beta_{ij}=-2.
\end{equation}
Plugging in \eq{INQ0} with 
\eq{xiQ} and \eq{chi4Q} in \eq{lau:Q}, we obtain an equation for 
$\inmu{N}{\m}{\bxi}$ similar to \eq{eq:I0} in terms of the quaternions
$\chi$. It is equivariant under $SL(2,\H)$. An invariant
set of equations is obtained by taking trace of the matrices involved. 
Taking trace on both sides the equations are expressed in terms of the 
quantities \eq{trchi}. We have, for each pair $(i,j)$,
\begin{equation}
	\label{eq:IQ0}
	\begin{split}
		&\sum_{A,B}
		\left(\sum_{\substack{k,l\\1\leq k,l\leq N\\
		k\neq i, l\neq j}}\alpha^A_{ik}\alpha^B_{jl}
		\tau_{ijkl}\right) 
	\xi_A\xi_B\pa_A\pa_B I_0(\bxi)\\
		&+\sum_A\left(4\alpha^A_{ij}
		+\sum_{\substack{k,l\\1\leq k,l\leq N\\k\neq i, l\neq j}}
		\left(\alpha^A_{ik}\alpha^A_{jl}
		+\alpha^A_{ik}\beta_{jl}
		+\alpha^A_{jl}\beta_{ik}\right)\tau_{ijkl}\right)
	\xi_A\pa_AI_0(\bxi)\\
		&+\left(4\beta_{ij}
		+\sum_{\substack{k,l\\1\leq k,l\leq N\\k\neq i, l\neq j}}
		\beta_{ik}\beta_{jl}\tau_{ijkl}\right)I_0(\bxi)=0,
	\end{split}
\end{equation}
which generalizes \eq{eq:I0}. In order to write the equations in terms of 
cross ratios we need to relate the trace and determinant of $\chi_{ijkl}$. 
To this end let us first note that
\begin{equation}
\begin{split} 
	\chi_{ijkl}\chi_{ijlk} &= 
	Q_{ij}Q_{ik}^{-1}Q_{kl}Q_{jl}^{-1} 
	\underline{Q_{ij}}Q_{il}^{-1}Q_{lk}Q_{jk}^{-1} \\
	&= Q_{ij}Q_{ik}^{-1}Q_{kl}Q_{jl}^{-1} 
	\underline{(Q_{il}-Q_{jl})}Q_{il}^{-1}Q_{lk}Q_{jk}^{-1} \\
	&= Q_{ij}Q_{ik}^{-1}Q_{kl}Q_{jl}^{-1}\underline{Q_{lk}}Q_{jk}^{-1} 
	- Q_{ij}Q_{ik}^{-1}\underline{Q_{kl}}Q_{il}^{-1}Q_{lk}Q_{jk}^{-1} \\
	&= Q_{ij}Q_{ik}^{-1}Q_{kl}Q_{jl}^{-1}\underline{(Q_{jk}-Q_{jl})}
            Q_{jk}^{-1} 
	- Q_{ij}Q_{ik}^{-1}\underline{(Q_{il}-Q_{ik})}Q_{il}^{-1}Q_{lk}
              Q_{jk}^{-1} \\
	&= \chi_{ijkl}+\chi_{ijlk},
\end{split}
\end{equation}
where the underlined terms indicate the replacements made at various 
intermediate stages.
Since $\chi_{ijkl}$ defined in \eq{chi4Q} is a quaternion, this is 
an equation of $2\times 2$ complex matrices. Let us rewrite it  as
\begin{equation}
	\label{chiid}
	\chi_{ijkl}=-(\mathbf{I}_2-\chi_{ijkl})\chi_{ijlk},
\end{equation}
where $\mathbf{I}_2$ denotes the $2\times 2$ identity matrix. 
We further note that for any $2\times 2$ matrix $M$ 
the identity
\begin{equation}
	\det (\mathbf{I}_2+ M)=1+\tr M+ \det M
\end{equation}
holds. 
Taking determinant of both sides of \eq{chiid} and using this identity
we derive
\begin{equation}
	\label{taudet}
	\tau_{ijkl}=1-|\chi_{lijk}|+|\chi_{ijkl}|.
\end{equation}
This relation will be used to express $\tau_{ijkl}$ in terms of the 
cross ratios in  equation \eq{eq:IQ0}.
\subsection{Four points}
Let us write down the invariant case of four points $N=4$ explicitly. Since the 
equations are rather cumbersome, we present the forms obtained
by choosing $\beta_{14}=\beta_{34}=0$, a freedom allowed by \eq{beta}.
We choose the independent cross ratios as in \eq{2inv}, namely,
\begin{equation}
	\label{xy:def}
	x=|\chi_{1234}|,\quad y=|\chi_{4123}|.
\end{equation}
In terms of these the equation
\eq{eq:IQ0} yields two equations for the invariant $I_0$, for $(i,j)=(1,2)$ 
and $(1,3)$, namely,
\begin{equation}
	\label{App:4}
	\begin{split}
		(x+y-1)\vt_{x}^2I_0(x,y)&+2x\vt_{xy}I_0(x,y)
		-\big(x(\beta_{13}+\beta_{24})+(1-y)\beta_{12}\big)
		\vt_xI_0(x,y)\\
		&\qquad+x(2+\beta_{12})\vt_yI_0(x,y)
		+x\beta_{13}\beta_{24}I_0(x,y)=0,\\
		(x+y-1)\vt_{y}^2I_0(x,y)&+2y\vt_{xy}I_0(x,y)
		-\big(y(\beta_{13}+\beta_{24})+(1-x)\beta_{23}\big)
		\vt_yI_0(x,y)\\
		&\qquad+y(2+\beta_{23})\vt_yI_0(x,y)
		+y\beta_{13}\beta_{24}I_0(x,y)=0,
	\end{split}
\end{equation}
where $\vt_x=x\tfrac{\pa}{\pa x}$ denotes the logarithmic derivative.
These are solved by the fourth Appell function \cite{chen,hyperdire},
$F_4$. The general solution is 
\begin{equation}
\begin{split}
\label{F4:sol}
I_0(x,y)=
&C_1F_4(2-\mu_2,\mu_4,\mu_3+\mu_4-1,\mu_1+\mu_4-1;x,y)+\\
&C_2x^{2-\mu_3-\mu_4}F_4(\mu_1,2-\mu_3,\mu_1+\mu_2-1,\mu_1+\mu_4-1;x,y)+\\
&C_3y^{2-\mu_1-\mu_4}F_4(\mu_3,2-\mu_1,\mu_3+\mu_4-1,\mu_2+\mu_3-1;x,y)+\\
&C_4x^{2-\mu_3-\mu_4}y^{2-\mu_1-\mu_4}F_4(2-
\mu_4,\mu_2,\mu_1+\mu_2-1,\mu_2+\mu_3-1;x,y),
\end{split}
\end{equation}
where $C_1,C_2,C_3,C_4$ are arbitrary constants and 
we used solutions of \eq{beta} with $\beta_{14}=\beta_{34}=0$
and $|\m|=4$.  Plugging in the four 
solutions in terms of this for $I_4(x,y)$, \eq{INQ0} gives the complete
expression for $\inmu{4}{\m}{\Q}$.
Equations ensuing from the other choices of the indices are either
not independent, as for $(i,j)=(1,4)$, for example, or related to it
by analytic continuation. 
\subsection{Correlation functions}
The correlation functions are related to the integrals $\inmu{N}{\m}{\Q}$ 
exactly as in the two-dimensional case, \eq{gn}, namely,
\begin{equation}
\label{gnQ}
G^{\Delta_1,\Delta_2,\cdots,\Delta_N}_N(\phi_1,\phi_2,\cdots\phi_N) =
\mathcal{F}\left(
\prod\limits_{\stackrel{i,j}{1\leq i<j\leq N}}
z_{ij}^{\ell_{ij}}
	I_N^{\boldsymbol\m}(\Q)\right), 
\end{equation} 
satisfying \eq{amd}. Here we use $\phi_i=\phi(Q_i)$.
Considerations same as before lead to the two and three point functions,
\begin{gather}
\label{G2:Q}
	G^{\Delta_1,\Delta_2}_2(\phi_1,\phi_2)
=\pi^2C_{\Delta_1}
\delta_{\Delta_1,\Delta_2}
\G{0}\frac{1}{|Q_{12}|^{\Delta_1}},\\
\label{G3:Q}
G_3^{(\Delta_1,\Delta_2,\Delta_3)}(\phi_1,\phi_2,\phi_3) =
C_{\Delta_1\Delta_2\Delta_3}
|Q_{12}|^{-\frac{1}{2}(\Delta_1+\Delta_2-\Delta_3)}
|Q_{13}|^{-\frac{1}{2}(\Delta_1+\Delta_3-\Delta_2)}
|Q_{23}|^{-\frac{1}{2}(\Delta_2+\Delta_3-\Delta_1)}.
\end{gather} 
Let us point out that while in the two-dimensional case we considered only
chiral fields, in here we consider a general scalar field although
we retain the same notation for the constants as in the two-dimensional 
case. The integrations are thus over the four-dimensional space rather than on
contours now. 

For the higher ones we need, once again, a projector.  The projector in four 
dimensions is given by \eq{proj} with 
\begin{equation}
	\label{PidQ}
	\Pid=
	\frac{1}{N_{\Delta}}\int\frac{\phi_{\Delta}(Q)
	\phi_{\Delta}(Q')}{|Q-Q'|^{4-\Delta}} d^4Q d^4Q',
\end{equation}
where the constant of normalization is given by
\begin{equation}
	\label{ND:4}
	N_{\Delta}=
\frac{{\pi^6C_{\Delta}\G{0}}
\G{\Delta-2}\G{2-\Delta}}{\G{\Delta}\G{4-\Delta}}
\end{equation}

The expressions for the correlation functions assume exactly the same form
as in two dimensions, with quaternions in the integrals in lieu of complex 
variables and the values of $N_{\Delta}$ changed to \eq{ND:4} and $I_0$ 
taken to be a solution of \eq{eq:IQ0}. 
For example, the four-dimensional four-point function is given with
such changes from \eq{d2G4} by
\begin{equation}
	\label{d4G4}
	\begin{split}
		G_4^{\Delta_1,\Delta_2,\Delta_3,\Delta_4}(\phi_1,\phi_2,
\phi_3,\phi_4)&= \sum_{\Delta}
              \frac{C_{\Delta_1\Delta_2\Delta}C_{\Delta,\Delta_3,\Delta_4}}{%
              N_{\Delta}} 
		|Q_{12}|^{-\tfrac{\Delta_1+\Delta_2-\Delta}{2}}
		|Q_{34}|^{-\tfrac{\Delta_3+\Delta_4+\Delta-4}{2}}\\
		&\qquad\times\inmu{4}{\left(
			\tfrac{\Delta_1+\Delta-\Delta_2}{2},
			\tfrac{\Delta_2+\Delta-\Delta_1}{2},
			\tfrac{4+\Delta_3-\Delta_4-\Delta}{2},
			\tfrac{4+\Delta_4-\Delta_3-\Delta}{2}\right)
			}{Q_1,Q_2,Q_3,Q_4},
	\end{split}
\end{equation}
where $N_{\Delta}$ is given by \eq{ND:4} and $I_4$ is given by \eq{INQ0} 
with $\beta_{14}=\beta_{34}=0$, 
$\beta_{12}=2-\Delta$,
$\beta_{13}=\tfrac{\Delta_2-\Delta_1+\Delta-4}{2}$, 
$\beta_{23}=\tfrac{\Delta_1+\Delta_4-\Delta_2-\Delta_3}{2}$, 
$\beta_{24}=\frac{\Delta_3-\Delta_4+\Delta-4}{2}$ 
and \eq{App:4} with $\xi_1=x$ and $\xi_2=y$.
\section{Discussion and Summary}
\label{sec:dis}
To summarize, in this article
we study $N$-point correlation functions of conformal primaries 
of conformal field theories in two- and four-dimensional 
Euclidean spaces. In the former case the conformal group
is $SL(2,\C)$, while in the latter case it is $SL(2,\H)$. We demonstrate the
semblance of the computations in the two cases. 

Instead of copies of 
the conformal compactification of the Euclidean space within the
light cone in two higher dimensions, we choose to work directly with the 
Fulton-Macpherson compactification of the $N$-point configuration space. 
For the four-dimensional Euclidean space with infinity adjoined,
$M=\R^4\cup\{\infty\}$, the  configuration space of $N$ points is
        \begin{equation}
                C_N(M)=M^N\setminus\{q_i\in M,\ q_i\neq q_j;\ i,j=1,2,\cdots,N\}
        \end{equation}
The Fulton-Macpherson completion is achieved by considering the embedding
\cite{kmv,sinha}
        \begin{equation}
                \begin{split}
                        &\gamma: C_N(M)\rt M^N\times
\big(S^3\big)^{\binom{N}{2}}
                \times[0,\infty]^{\binom{p}{3}},\\
                        \big(q_1,q_2,\cdots,q_N\big)\longmapsto
&\big(q_1,q_2,\cdots,q_N,v_{12},\cdots,v_{(N-1)N},a_{123},
                        \cdots,a_{(N-2)(N-1)N}\big),
                \end{split}
        \end{equation}
where each of 
\begin{equation} 
v_{ij}=(q_i-q_j)/|q_i-q_j|
\end{equation} 
describes a three-sphere $S^3$ and the scalars
\begin{equation} 
a_{ijk}=\frac{|q_{ij}|}{|q_{ik}|}
\end{equation} 
assume values in the non-negative real line. 
Representations of the conformal group, in particular, the integral
$I^{\m}_N$, is to be chosen from among the 
functions of these variables. Invariance under translation bars a
representation to depend on $q_i$ alone and rotational invariance keeps it
from having dependence on $v_{ij}$. 
The difference $|q_i-q_j|$, however, is allowed. 
Let us note that $v_{ij}$ will appear in the correlation functions of
higher rank tensor fields.  
The expression \eq{INQ0} is thus a regular function on the Fulton-Macpherson
compactification of the configuration space with the cross-ratios expressed
as
\begin{equation}
|\chi_{ijkl}|=a_{ijk}a_{lkj}.
\end{equation}
Correlation functions are given
by representations of the conformal groups on the configuration space. We
obtain the representations of the groups directly without recourse to the
corresponding Lie algebras. Consistency of the expressions can be verified
by writing down the generators of the groups as differential operators. The
integrals then get related to conformal blocks.
While using the conformal algebra is effective in two
dimensions, non-commutativity of the quaternions render the computations
difficult in the four dimensional case. This approach also avoids building
the $N$-point functions from the ``single-particle" representations by tensoring
and eschews the insertion of ``single-particle" Casimirs. 

In two dimensions, the representation of the conformal or the M\"obius group
is obtained in terms of a Lauricella system. A differential equation for the
invariant part is derived for $N>3$ from the Lauricella system. 
We present solutions for $N=4,5$,
the former in terms of Gauss Hypergeometric function and the latter in terms
of the Appell function $F_2$. Parsing the correlation functions into 
three-point functions by inserting projectors we write integral formulas for
the correlation functions from the representations. The projectors themselves
are expressed in terms of the two-point Lauricella system. 

These considerations directly generalize to the four-dimensional case. We
define integrals in terms of determinants of quaternions. Differentiating
with the complex $2\times 2$ matrices representing quaternions we then set up a
generalized Lauricella system of differential equations for the integrals.
Representations of the conformal group $SL(2,\H)$ are  then obtained from
the solutions of the differential equation. The invariant cross-ratios are
given by the determinant of  quaternions. 
In order to write the equations
for the invariant part we use the relation between the trace and determinant
of $2\times 2$ matrices. While the equations are obtained for an arbitrary
$N$, we present the computation for $N=4$, in which case the integral is
given by the Appell function $F_4$. As in the two dimensional case, the
correlation functions are parsed using projectors obtained as solutions to
the Lauricella system for $N=2$, without requiring it to transform under the
conformal group. Let us stress that the correlations functions in the
two-dimensional case have been known for decades. The four-dimensional
four-point function in the comb channel has been worked out earlier
\cite{rosenhaus} and our results match these expressions. It is their direct
connection with the Fulton-Macpherson compactification of configuration
spaces of $N$ marked points and the quaternionic Lauricella system that 
governs them in four dimensions that is novel in here.  

Let us also point out that the projectors \eq{PidN} and \eq{PidQ} can be
expressed in terms of the so-called \emph{shadow operator} by choosing to
perform the integration over $y$ first \cite{sd}. We have chosen to postpone
it to a later stage of the computation in order to relate to the integrals
$I^{\m}_N$. Further, we have presented the most general expressions for
the solutions of the Lauricella systems. However, the correlation functions
were parsed in terms of three-point functions. In order to be concomitant
with the operator product expansion some of the terms must be discarded 
in the final expressions for the correlation functions by
the monodromy projection \cite{sd}. For example, only one of the two terms in
\eq{I0:solgen} is to be retained in \eq{d2G4},  namely,
\begin{equation}
\begin{split}
&\left.\inmu{4}{\left(
			\tfrac{2-\Delta+\Delta_1-\Delta_2}{2},
			\tfrac{2-\Delta+\Delta_2-\Delta_1}{2},
			\tfrac{\Delta+\Delta_3-\Delta_4}{2},
			\tfrac{\Delta+\Delta_4-\Delta_3}{2}\right)
			}{z_1,z_2,z_3,z_4}\right|_{\text{Projected}}\\
			&=z_{12}^{\Delta-1}z_{13}^\frac{\Delta_2-\Delta_1-\Delta}{2}z_{23}^\frac{\Delta_1-\Delta_2-\Delta_3+\Delta_4}{2}z_{24}^\frac{\Delta_3-\Delta_4-\Delta}{2}F\left(\tfrac{\Delta+\Delta_1-\Delta_2}{2},\tfrac{\Delta+\Delta_4-\Delta_3}{2},\Delta;\xi\right).
\end{split}
\end{equation}
Similarly, only two of the four terms in \eq{F4:sol} survive the monodromy
projection. The integral $I_4$ to be used in \eq{d4G4} is 
\begin{equation}
\label{I4P}
\begin{split}
&\left.{\inmu{4}{\left(
			\tfrac{\Delta_1+\Delta-\Delta_2}{2},
			\tfrac{\Delta_2+\Delta-\Delta_1}{2},
			\tfrac{4+\Delta_3-\Delta_4-\Delta}{2},
			\tfrac{4+\Delta_4-\Delta_3-\Delta}{2}\right)
			}{Q_1,Q_2,Q_3,Q_4}}\right|_\text{Projected}\\
			&\quad = 
x^{\Delta-2}|Q_{12}|^{2-\Delta}|Q_{13}|^{\tfrac{\Delta_2-\Delta_1+\Delta-2}{2}}|Q_{23}|^{\tfrac{\Delta_1-\Delta_2-\Delta_3+\Delta_4}{2}}|Q_{24}|^{\tfrac{\Delta_3-\Delta_4+\Delta-4}{2}}\times\\
			&\qquad\Big(C_2F_4\left(\tfrac{\Delta+\Delta_1-\Delta_2}{2},
\tfrac{\Delta+\Delta_4-\Delta_3}{2},\Delta-1,\tfrac{\Delta_1-\Delta_2-\Delta_3
+\Delta_4}{2}+1;x,y\right)\\
			&\qquad+C_4y^{\tfrac{-\Delta_1+\Delta_2+\Delta_3-\Delta_4}{2}}F_4\left(\tfrac{\Delta+\Delta_3-\Delta_4}{2},\tfrac{\Delta+\Delta_2-\Delta_1}{2},\Delta-1,\tfrac{\Delta_2-\Delta_1+\Delta_3-\Delta_4}{2}+1;x,y
\right)\Big).
\end{split}
\end{equation}
In all the cases the monodromy considerations project out part of the basis
of the Lauricella system. 
The integrals entering the expressions for correlation functions are
generically multi-valued, rendering their direct evaluation complicated.
Expressing these as solutions to  
differential equations  may be very useful in this
regard. The situation is similar to the evaluation of periods of algebraic
varieties, whose evaluation in various domains of convergence
is substantially facilitated by expressing 
them as solutions Picard-Fuchs differential equations.
The Lauricella system developed here in terms of quaternions are quite
general. We expect this formalism to be useful in computing the correlation
functions in four dimensions as well as in computing Feynman integrals in 
quantum field theories. 
\section*{Acknowledgement}
KR thanks Dileep Jatkar for sharing his insight. 

\end{document}